# Time before Time

## Classifications of universes in contemporary cosmology, and how to avoid the antinomy of the beginning and eternity of the world

Rüdiger Vaas

Center for Philosophy and Foundations of Science, University of Gießen
Otto-Behaghel-Str. 10 C, D – 35394 Gießen, Germany

Ruediger.Vaas@t-online.de

**Abstract**

Did the universe have a beginning or does it exist forever, i.e. is it eternal at least in relation to the past? This fundamental question was a main topic in ancient philosophy of nature and the Middle Ages. Philosophically it was more or less banished then by Immanuel Kant's *Critique of Pure Reason*. But it used to have and still has its revival in modern physical cosmology both in the controversy between the big bang and steady state models some decades ago and in the contemporary attempts to explain the big bang within a quantum cosmological framework. This paper has two main goals: First a conceptual clarification and distinction of different notions of „big bang" and „universe" is suggested, as well as a multiverse taxonomy and a classification of initial and eternal cosmologies. Second, and with the help of this analysis, it is shown how a conceptual and perhaps physical solution of the temporal aspect of Immanuel Kant's „first antinomy of pure reason" is possible, i.e. how our universe in some respect could have *both* a beginning *and* an eternal existence. Therefore, paradoxically, there might have been a time before time or a beginning *of* time *in* time.

__________________________________________________________________________________

# 1. A universe with or without a beginning

Although modern physical cosmology has been able to emancipate itself considerably from its philosophical predecessors, it is still occupied with some of their fundamental questions (cf. Kanitscheider 1991). One of them is the problem of the finiteness versus infiniteness of time and space. The philosophical implications of current scientific approaches to these problems and the big bang, should they turn out to be true, are far-reaching; and they are based on many (partly speculative) premises as well as concepts, which are not always sufficiently clear in scientific practice and popularization. Therefore these approaches are also an interesting subject of reflections for philosophers of science (cf. Bartels 1996).

Immanuel Kant, in his *Critique of Pure Reason* (1781/1787), argued that it is possible to prove *both* that the world has a beginning *and* that it is eternal (first antinomy of pure reason, A426f/B454f). As Kant believed he could overcome this „self-contradiction of reason" („Widerspruch der Vernunft mit ihr selbst", A740) by what he called „transcendental idealism", the question whether the cosmos exists forever or not has almost vanished in philosophical discussions. This is somewhat surprising, because Kant's argument is quite problematic (cf., e.g., Heimsoeth 1960, Wilkerson 1976, Wike 1982, Smith 1985, Schmucker 1990, Kanitscheider 1991, pp. 439 ff, Malzkorn 1999, Falkenburg 2000). In the twentieth century, however, the question became once again vital in the context of natural science, culminating in the controversy between big bang and steady state models in modern physical cosmology (Kragh 1996). In recent years, it has reappeared in the framework of quantum cosmology (Vaas 2001d & 2002a), where, on the one hand, there are instanton models that assume an *absolute beginning of time* (e.g. Vilenkin 1982 & 1984, Hawking & Hartle 1983, Hawking & Turok 1998, Turok & Hawking 1998), while other scenarios suppose that the big bang of our universe was only a *transition from an earlier state* (e.g. Linde 1983 & 1994 & 2003a & 2004, Blome & Priester 1991, Khoury *et al.* 2001 & 2003 & 2004, Steinhardt & Turok 2002abc, Turok, Malcolm & Steinhardt 2004, Turok & Steinhardt 2004ab), and that there are perhaps infinitely many such events.

General relativity breaks down at very small spatio-temporal scales and high energy densities. This is why quantum cosmology is needed. But in contrast to the framework of general relativity, which is theoretically well understood and has been empirically confirmed quite marvelously, the current approaches in quantum cosmology, string theory, etc., are still quite speculative, controversial, and almost without any empirical footing yet. Although it is on a much more sophisticated and abstract level, this situation somewhat resembles the pre-Socratic discussions of natural philosophy. This is a further reason why conceptual analysis and philosophical investigations of assumptions and implications in general might be useful here – both within and beyond physics.

This paper has two goals: First, some conceptual clarifications shall be made helping to classify different cosmologies and avoid confusion: They are about distinct meanings of the terms „big bang" and „universe" and offer a proposal for a multiverse taxonomy. Second, a speculative solution of Kant's „first antinomy of pure reason" within a framework of metaphysical realism and based on a distinction between two kinds of time (a microscopic and a macroscopic time scale) is suggested, which is compatible with some modern cosmological scenarios. Our universe in some respect could have *both* a beginning *and* an eternal existence. Therefore, paradoxically, there might have been a time before time or a beginning *of* time *in* time.

## 2. Different notions of „big bang", „universe", and „multiverse"

„Big bang" is an ambiguous term, which has led to many misunderstandings and prejudice. One should draw a distinction between at least four logically different meanings:
(1) the hot, dense early phase of our universe where the light elements were formed,
(2) the initial singularity,
(3) an absolute beginning of space, time, and energy, and
(4) the beginning of our universe, i.e. its elementary particles, vacuum state, and perhaps its (local) space-time.

That our universe originated from a big bang in the sense of (1) is almost uncontroversial. (2) is the relativistic cosmology's limit of backward extrapolation where the known laws of physics break down. Different models of quantum and string cosmology try to overcome this limit, and (3) and (4) classify their different scenarios. Those characterized by (3) might be called *initial cosmologies*. They postulate a very first moment (cf. Grünbaum 1991, Smith 2002) – at least if time is ultimately physically discrete, i.e. not continuously divisible. Those characterized by (4) are *eternal cosmologies*. There are different kinds of them, both in ancient and in modern cosmology: *static* ones (without irreversible changes on a coarse-grained level), *evolutionary* ones (with cumulative change), and *revolutionary* ones (with sharp phase-transitions). And they could have either a *linear* or a *cyclic* time. Option (4) also allows the possibility that our universe neither exists eternally, nor that it came into being out of nothing or out of a timeless state, but that space and time are not fundamental and irreducible at all, or that there was a time „before" the big bang – „big bang" in the sense of (1) –, as well as that there are other universes.

The term „universe" (or „world"), as it is used today, is also ambiguous. There are many different meanings of „universe", especially:
(1) everything (physically) in existence, ever, anywhere;
(2) the observable region we inhabit (the Hubble volume, roughly 27 billion light years in diameter), plus everything that has interacted (for example due to a common origin) or will ever or at least in the next few billion years interact with this region;
(3) any gigantic system of causally interacting things that is wholly (or to a very large extent or for a long time) isolated from others; sometimes such a locally causally connected collection is called a *multi-domain universe*, consisting of the ensemble of all sub-regions of a larger connected spacetime, the „universe as a whole", and this is opposed to multiverse in a stronger sense, i.e. the set of genuinely diconnected universes, which are not causally related at all.
(4) any system that *might* well have become gigantic, etc., even if it does in fact recollapse while it is still very small;
(5) other branches of the wavefunction (if it never collapses, cf. Wheeler & Zurek 1983, Barrett 1999, Vaas 2001e & 2004c) in unitary quantum physics, i.e. different histories of the universe (e.g. Gell-Mann & Hartle 1990 & 1993) or different classical worlds which are in superposition (e.g. DeWitt & Graham 1973);
(6) completely disconnected systems consisting of universes in one of the former meanings, which do or do not share the same boundary conditions, constants, parameters, vacuum states, effective low-energy laws, or even fundamental laws, e.g. different physically realized mathematical structures (cf. Tegmark 2004).
Nowadays, „multiverse" (or „world" as a whole) might be used to refer to everything in existence, while the term „universe" permits to talk of several universes (worlds) within the multiverse. In principle, these universes – mostly conceived in the meaning of (2), (3), or (4) – might or might not be spatially, temporally, dimensionally, and/or mathematically separated from each other (see table 1). Thus, there are not necessarily sharp boundaries between them.

*Table 1:*

## **A multiversal taxonomy**

| separation | aspects of separation | examples |
|---|---|---|
| spatio-temporal | • spatial (*see:* causal) | |
| | – exclusive | different quantum universes, eternal inflation |
| | – inclusive | embedment: universes in atoms (c.f. Blaise Pascal) or black holes; cf. an infinite universe in a finite quantum fluctuation |
| | • temporal | Phoenix universe, cyclic universe, recycling universe |
| | • dimensional | |
| | – strict | tachyon universe? |
| | – intermittent | flatland, brane-worlds |
| | – abstract | superspace, where universes are like leaves in a pile of paper |
| causal | • strict | |
| | – without a common generator | different universes or multiverses in the instanton, big bounce, soft bang, self-creation scenarios, different „bundles" with cosmic inflation |
| | – genealogical | splitting off (chaotic inflation, Lee Smolin's cosmic Darwinism) |
| | | many worlds/histories of quantum physics (if they do not interact at all) |
| | • continuous | *(e.g. due to a growing cosmic horizon)* |
| | – always | infinite space ($k = 0/-1$), eternal inflation, infinite branes |
| | – once | previously disconnected parts due to inflation, now in interaction again |
| | – coming | due to accelerated expansion because of dark energy |
| modal | • potential | *(therefore not every possibility would be realized!)* |
| | • real | modal realism (David Lewis); physical vs. metaphysical vs. logical |
| mathematical | • structural/axiomatic | Platonism, mathematical democracy, ultimate ensemble (Max Tegmark) |

(Note that these aspects of seperation are not necessarily excluding each other. Some multiverse kinds fit in several categories, i.e. chaotic inflation which describes universes which are spatially exclusive, but not dimensionally separated, causally future-strict, but not past-strict separated because they share a common mechanism which generates their existence.)

A distinction which follows naturally from the above might be called the unique-universe versus multiverse accounts:

• *Unique-universe accounts* view our universe as the only one (or at least the only one ever relevant for and subject of cosmological explanations and theories). It might have had predecessors (before the big bang) and/or successors (after a big crunch), but then the whole series can be taken as one single universe with spatio-temporal phase-transitions (e.g. as a revolutionary eternal cosmology). From the perspective of physical simplicity, epistemology and philosophy of science it is favorable to try to explain as much as possible with a unique-universe account, i.e. searching for a Theory of Everything with just one self-consistent solution that represents (or predicts) our universe. (Of course one could always argue that there are other, causally strictly separated universes too, which do not even share a common generator or a meta-law; but then they do not have any explanatory power at all and the claims for their existence cannot be motivated in any scientific useful way.)

- *Multiverse accounts*, however, assume the existence of other universes as defined above and characterized with respect to their separation in table 1. This is no longer viewed as mad metaphysical speculation beyond empirical rationality. There are many different multiverse accounts (see, e.g., Linde 1994, Smolin 1997, Deutsch 1997, Vaas 1998 & 2004e, Rees 2001, Gouts 2003, Tegmark 2004, Davies 2004, Carr 2005) and even some attempts to classify them quantitatively (see, e.g., Deutsch 2001 for many worlds in quantum physics, and Ellis, Kirchner & Stoeger 2004 for physical cosmology).

It is almost an ingredient of standard big bang cosmology to assume *cosmic inflation* – and there is already some observational evidence (see Spergel et al. 2003, Peiris et al. 2003, Vaas 2003b) although it might be explained otherwise (cf. Khoury, Steinhardt & Turok 2003, Turok & Steinhardt 2004a). According to this scenario, almost immediately after the big bang our universe underwent a – perhaps only very short – epoch of exponential expansion (see, e.g., Abbott & Pi 1986, Blau & Guth 1987, Linde 1994 & 2004, Lidsey et al. 1997, Guth 1997 & 2000ab & 2001 & 2002, Vaas 2001b). This superluminal increase of space has a doubling time in the order of $10^{-37}$ seconds, and there were at least a hundred of these doublings. This is the reason why our entire observable universe is so extremely homogeneous on large scales. After the inflationary epoch, which was driven by a new physical field called the inflaton (or a combination of different fields or geometric properties of space itself), the energy of that field was released and ultimately converted to matter and radiation. Models of cosmic inflation can solve deep physical and cosmological problems which need not be discussed here. Inflation is a property of a certain kind peculiar repulsive-gravity material with negative pressure or a special kind of vacuum, called the false vacuum (both terms, „material" and „vacuum", are somewhat misleading, and they are instantiated by the inflaton which is fundamentally unstable). If and where inflation stops, thermalized regions form – large bubbles of „true" vacuum, sometimes called pocket universes. Our universe is a tiny part of such a much more slowly expanding bubble. But these bubbles are still surrounded by the false vacuum. In fact, this false vacuum never decays completely because the expansion is much faster than the decay. Therefore, inflation never stops. This is the basic idea of eternal inflation. More precisely we should say: future-eternal inflation, because it is a matter of controversy whether inflation is also past-eternal or not (see below). – Inflation is not the only argument for multiverse accounts. Also most theories of quantum cosmology (e.g. based on string theory) assume or cannot avoid the existence of different universes (Susskind 2003 & 2004a, Douglas 2003 & 2004). To explain the existence and properties of our universe it would be favorable to find arguments showing that universes of our kind (or similar kinds) are more probable than other ones or that our universe is typical in some sense. If this is not possible – if, e.g., all kinds of universes are equally probable or our universe even has a very low probability of coming into being compared to others –, one has to adopt observational selection effects, also known as the weak anthropic principle (see Barrow & Tipler 1986 and Vaas 2004e for an extensive review).

Though the multiverse idea is quite speculative and comes along with deep scientific, epistemological and ontological problems, it has some explanatory power and is, at least in some variants, probably here to stay. „The idea of a multiverse – an ensemble of universes or universe domains – has received increasing attention in cosmology, both as the outcome of the originating process that generated our own universe, and as an explanation for why our universe appears to be fine-tuned for life and consciousness" (Ellis, Kirchner & Stoeger 2004, p. 921). „Attempts to sharpen the discussion and provide a more rigorous treatment of concepts such as the number of universes, the probability measures in parameter space, and objective definitions of infinite sets of universes, have not progressed far. Nevertheless, the multiverse idea has probably earned a permanent place in physical science, and as new physical theories are considered in the future, it is likely that their consequences for biophilicity and multiple cosmic regions will be eagerly assessed" (Davies 2004, online p. 14).

One might call the whole set of different universes the multiverse. But it could be true that there are even different sets of totally spatio-temporally and causally strictly separated multiverses, e.g. different bunches of chaotically inflating, or self-created multiverses (see below). In that case it remains meaningful to have a term with a still broader extension, namely *cosmos*. So „cosmos" shall be taken as the all-embracing term for everything in existence which might or might not be the set of different multiverses, while a (or the) multiverse consists of different universes.

## 3. Initial and eternal cosmologies

Historically, both cosmologies with and without a beginning were common. This fundamental question was a main topic in ancient philosophy of nature. It culminated in the Middle Ages. But philosophically it was more or less banished then by Immanuel Kant's *Critique of Pure Reason*. Some historical philosophical views and their influential proponents are summarized in table 2.

*Table 2:*

**Classical initial and eternal cosmologies**

| possibilities (examples) | prominent proponents (beyond mythology) |
|---|---|
| **eternal ur-substance**:  transformations, aggregations,  or recurrence | Thales, Anaximander, Xenophanes, Empedocles, Anaxagoras, Leucippus, Demokritus, Heraclitus, Parmenides, Plato, Stoa |
| **eternity: no beginning and no transience/destructibility**  of matter, motion,  the celestial sphere,  or deity | Aristotle; Averroes (Abù al-Walìd Muhammad b. Ahmad Ibn Rushd); *as a possibility:* Thomas Aquinas, Siger of Brabant, Boetius of Dacia, Wilhelm Ockham, Giles of Rome, Godfrey of Fontaines, Henry of Harclay, Thomas of Wylton, Thomas of Strassburg; Pietro D'Abano, Johannes Jandunus, Pietro Pomponazzi; Giordano Bruno, Baruch de Spinoza |
| **eternity of the world, but emanation of the deity** | Plotinus, Proclus; Origen; Abu Nasr Al-Faràbi (Alfarabius), Abù'Alì al-Husayn Ibn Sinna (Avicenna) |
| **timlessness of creation** | Johannes Scotus Eurigena, Meister Eckhart |
| *creatio continua*  continuous maintenance and  re-creation of the world | Anaximenes; Stoa, Cicero; Augustine of Hippo; Gregor the Great, Thomas Aquinas, Henry of Gent, Petrus Aureoli; René Descartes |
| **absolute beginning, creation,** *creatio ex nihilo* | Johannes Philoponos, Augustine of Hippo; Yaaqub ibn Ishaq Al-Kindi (Alkindus); Hugo & Richard & Bernard of St. Viktor, Bernard & Thierry of Chartres, Wilhelm of Conches, Clarenbaldus of Arras; Abù Hamìd Muhammad Al-Ghazàli, Moses Maimonides; Albertus Magnus, Wilhelm of Auvergne, Robert Grosseteste, Roger Bacon, Bonaventura, Matthew of Aquasparta, Henry of Gent, John Pecham, Richard of Middleton, Peter Aureol, William of Alnwick, Henry Totting of Oyta, Marsilius of Inghen |
| **beginning, no end** | Philo of Alexandria |
| **transcendental appearance** | Immanuel Kant |

(Note that not all those possibilites are necessarily in contradiction in every respect. A *creatio ex nihilo* and a *creatio continua*, for instance, could both be true.)

In modern physical cosmology, the situation is secularized, but even more complex, because general relativity and quantum cosmology offer additional possibilities. An overview is given in figure 1 and table 3.

*Figure 1:*

**Spacetime diagrams of different cosmological models**

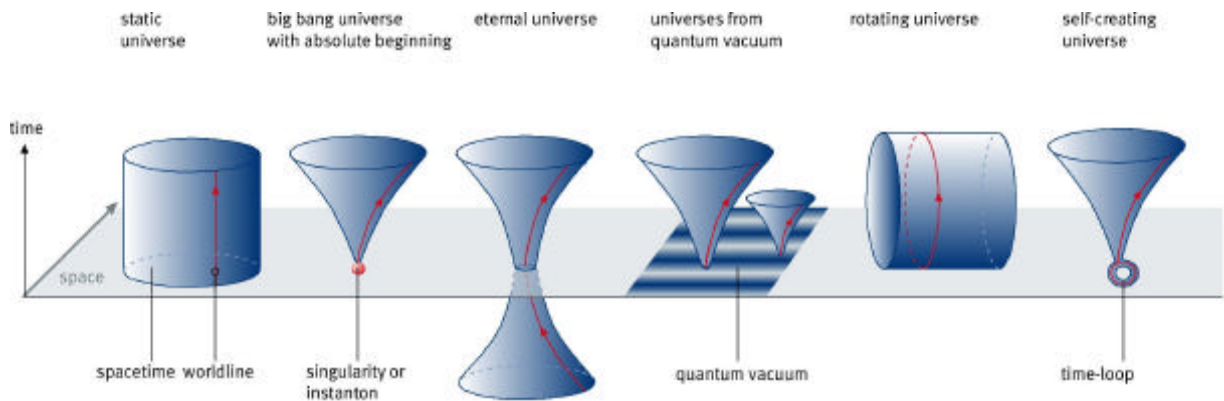

*Table 3:*

## Modern initial and eternal cosmologies

| possibilities | models (space-time) and their main proponents |
|---|---|
| *beginning and an end* | • **classical big bang/big crunch**: Alexander Friedmann (1922), Stephen Hawking & Roger Penrose (1965 ff)<br>• **quantum tunnel effect**: Alexander Vilenkin (1982 ff)<br>• **no boundary instanton**: Stephen Hawking & James Hartle (1983) |
| *beginning, but no end* | • **classical big bang/big whimper**: Alexander Friedmann (1924), Georges Lemaître (1927), Stephen Hawking & Roger Penrose (1965 ff)<br>• **phoenix universe** (global!): Georges Lemaître (1933), Richard C. Tolman (1934)<br>• **quantum tunnel effect and eternal inflation**: Alexander Vilenkin (1982 ff)<br>• **cosmic Darwinism:** Lee Smolin (1992 ff)<br>• **no boundary instanton**: Stephen Hawking & Neil Turok (1998) |
| *no beginning and no end (static. vs. evolutionary vs. revolutionary)* | • **static universe**: Albert Einstein (1917)<br>• **empty expanding universe**: Willem de Sitter (1917)<br>• **eternal expansion out of a static universe**: Arthur S. Eddington (1930)<br>• **steady state**: Hermann Bondi, Thomas Gold & Fred Hoyle (1948 ff)<br>• **quasi-steady state**: Fred Hoyle, Geoffrey Burbidge & Jayant V. Narlikar (1993 ff)<br>• **chaotic inflation** (global!): Andrei Linde (1983 ff)<br>• **Planckian cosmic egg** (global!): Mark Israelit & Nathan Rosen (1989 ff)<br>• **big bounce**: Hans-Joachim Blome & Wolfgang Priester (1991)<br>• **ekpyrotic and cyclic universe** (global!): Paul Steinhardt & Neil Turok et al. (2001 ff) |
| *no beginning, but an end* | • **collapse out of a static universe**: Arthur S. Eddington (1930) |
| *cycle (recurrence)* | • **oscillating universe** (local!): Mark Israelit & Nathan Rosen (1989 ff), Redouane Fakir (1998)<br>• **cyclic universe** (local!): Paul Steinhardt & Neil Turok et al. (2002 ff)<br>• **circular time in a rotating universe**: Kurt Gödel (1949 ff)<br>• **big brunch/time-reversal**: Claus Kiefer & H. Dieter Zeh (1995) |
| *time-loop with/without end* | • **self-creating universe**: John Richard Gott III & Li-Xin Li (1998) |
| *pseudo-beginning with/without a local end* | *background-dependent:*<br>• **soft bang/emergent universe**: Eckard Rebhan (2000), George F. R. Ellis & Roy Maartens et al. (2003)<br>• **quantum fluctuation, de Sitter instability etc.**: Edward Tryon (1973), Robert Brout et al. (1978 ff), Alexei A. Starobinsky (1979 ff), David Atkatz & Heinz R. Pagels (1982), John Richard Gott III (1982), Mark Israelit (2002)<br>• **pre-big bang**: Gabriele Veneziano & Maurizio Gasperini (1991 ff)<br>*background-independent:*<br>• **pregeometry**: John A. Wheeler (1975), Peter W. Atkins (1981), Stephen Wolfram (2002)<br>• **loop quantum cosmology**: Abhay Ashtekar & Martin Bojowald et al. (2002 ff) |

Although eternal cosmologies might appear to be more naturally related to multiverse accounts, the temporal and quasi-spatial issues, i.e. initial/eternal cosmologies and unique universe/multiverse accounts, are logically independent. Every possible combinations of those features were suggested and defended, as the following examples show:

*Table 4:*

|  | unique-universe approach | multiverse approach |
|---|---|---|
| **initial cosmologies** | classical singularity big bang model<br>Hawking-Hartle/Turok instantons | Vilenkin's quantum tunneling out of nothing |
| **past-eternal cosmologies** | steady state model<br>Eddington's past-static universe | chaotic inflation<br>ekpyrotic and cyclic universes |

Of course classifications like those in table 3 have crucial limitations. They are not meant to put the different models rigidly into different categories or placing them onto the bed of Procrustes, but to get a simplified overview of these complex issues as well as to highlight and contrast the differences. Although such a framework or taxonomy is useful for systematic, conceptual and didactic reasons, it is important to be aware of the simplifications and controversial ambiguities of it. Before discussing some of them (see chapter 6 below), another important distinction is necessary, however – that between microscopic and macroscopic time. It is also crucial to understand a third option between initial and eternal cosmologies: the pseudo-beginning of universes.

Both initial and eternal cosmologies have severe explanatory problems, so it is useful to search for alternatives.

Eternal cosmologies need not assume a first cause or accident, but they shift the burden of explanation into the infinite past. Although every event might be explicable by earlier events and causal laws, eternal cosmologies cannot even address why a temporally infinite cosmos exists and why it is the way it is. And there might be even deeper problems: Because we are able to assign a symbol to represent ‚infinity' and can manipulate such a symbol according to specified rules, one might assume that corresponding infinite entities (e.g. particles or universes) exist. But the actual (i.e. realized in contrast to potential or conceptual) physical (in contrast to the mathematical) infinite has been criticized vehemently being not constructible, implying contradictions etc. (cf. Hilbert 1964, p. 136-151, Spitzer 2000, Stoeger, Ellis & Kirchner 2004, ch. 5). If this would be correct it should also apply to an infinite past. (A future-eternal cosmos might be less problematic if it is viewed as an unfolding, unbounded i.e. only potential one.) This is an controversial issue, but it might be seen at least as another motivation to search for alternatives to past-eternal cosmologies.

Initial cosmologies, on the other hand, run into deep metaphysical troubles to explain how something could come out of nothing and why is there something rather than nothing at all (cf. Nozick 1981). Even the theological doctrine of *creatio ex nihilo* does not start with nothing at all but with something, that is God, so the principle „ex nihilo nihil fit" still holds. And contemporary secularized *ex-nihilo* initial cosmologies usually claim, as Alexander Vilenkin said (quoted in Vaas 2003c, p. 45), that there were at least the laws of physics even if there was nothing more at all. (Concerning his own model, Vilenkin (1982, p. 26) admitted that „The concept of the universe being created from nothing is a crazy one", and his analogy with particle pair creation only deepens the problem, because matter-antimatter particles do not pop out of nothing but are transformations of energy which is already there.) Similarly, Heinz Pagels (1985, p. 347) subscribed to some kind of platonism with respect to physical laws: „This unthinkable void converts itself into the plenum of existence – a necessary consequence of physical laws. Where are these laws written into that void? What ‚tells' the void that it is pregnant with a possible universe? It would seem that even the void is subject to law, a logic that exists prior to space and time." And Stephen Hawking (1988, p. 174) asked „Even if there is only one possible unified theory, it is just a set of rules and equations. What is it that breathes fire into the equations and makes a universe for them to describe? The usual approach of science of constructing a mathematical model cannot answer the question of why there should be a universe for the model to describe. Why does the universe go to all the bother of existing?" But if one does not subscribe to an origin of something (or everything) from really nothing, one need not accept platonism with respect to the ontological status of physical laws (Vaas 2003c). They might simply be seen as the outcome of invariant properties of nature. If so, they do not govern nature but are instantiated from it. They are abstract descriptions – like a model or theory of reality is not to be confused with reality itself.

So from what has been sketched here, it is helpful to search for a „third way" between initial and eternal cosmologies to explain more than the latter but not fall into the problems of the former. Pseudo-beginning cosmologies, as they should be called here, might offer such a middle course.

## 4. A solution for the antinomy of the beginning and eternity of the world

Kant's first antinomy makes the error of the excluded third option, i.e. it is not impossible that the universe could have *both* a beginning *and* an eternal past. If some kind of metaphysical realism is true, including an observer-independent and relational time, then a solution of the antinomy is conceivable. It is based on the *distinction between a microscopic and a macroscopic time scale*. Only the latter is characterized by an asymmetry of nature under a reversal of time, i.e. the property of having a global (coarse-grained) evolution – an arrow of time (Zeh 2001, Vaas 2002c, Albrecht 2003) – or many arrows, if they are independent from each other. (Note that some might prefer to speak of an arrow *in* time, but that should not matter here.) Thus, the macroscopic scale is by definition temporally directed – otherwise it would not exist. (It shall not be discussed here whether such an arrow must be observable in principle, which would raise difficult questions, e.g. in relation to an empty, but globally expanding universe.)

On the microscopic scale, however, only local, statistically distributed events *without* dynamical trends, i.e. a global time-evolution or an increase of entropy density, exist. This is the case if one or both of the following conditions are satisfied: First, if the system is in thermodynamic equilibrium (e.g. there is a huge number – or degeneracy – of microscopic states identifiable with the *same* coarse-grained state). And/or second, if the system is in an extremely simple ground state or meta-stable state. (Meta-stable states have a local, but not a global minimum in their potential landscape and, hence, they can decay; ground states might also change due to quantum uncertainty, i.e. due to local tunneling events.) Some still speculative theories of quantum gravity permit the assumption of such a global, macroscopically time-less ground state (e.g. quantum or string vacuum, spin networks, twistors). Due to accidental fluctuations, which exceed a certain threshold value, universes can emerge out of that state. Due to some also speculative physical mechanism (like cosmic inflation) they acquire – and, thus, are characterized by – directed non-equilibrium dynamics, specific initial conditions, and, hence, an arrow of time. (It could be defined, for instance, by the cosmic expansion parameter or by the increase of entropy.) Note that, strictly speaking, such universes are not „inside" or „embedded in" the vacuum ground state but cut their cords and exist in some respects „anywhere else".

It is a matter of debate (cf., e.g., Price 1996, Vaas 2002c) whether such an arrow of time is
1) irreducible, i.e. an essential property of time (e.g. Maudlin 2002),
2) governed by some unknown fundamental and not only phenomenological law (e.g. Penrose 1989, Prigogine 1979),
3) the effect of specific initial conditions (cf. Albrecht 2004, Schulman 1997, Zeh 2001) or
4) of consciousness (if time is in some sense subjective, e.g. Kant 1781/1787), or
5) even an illusion (e.g. Barbour 2000).

Many physicists favour special initial conditions, though there is no consensus about their nature and form. But in the context at issue it is sufficient to note that such a macroscopic global time-direction is the main ingredient of Kant's first antinomy, for the question is whether this arrow has a beginning or not.

(A brief comment might be appropriate here: Of course, *if* – and this is a big IF – time's arrow is inevitably subjective, ontologically irreducible, fundamental and not only a kind of illusion, thus if some form of metaphysical idealism for instance is true, *then* both physical cosmology and the proposal of this paper as well as any reasoning about a time before time is mistaken or quite irrelevant. However, if we do not want to neglect an observer-independent physical reality and adopt solipsism or other forms of idealism – and there are strong arguments in favor of some form of metaphysical realism –, Kant's rejection seems hasty. Furthermore, if a Kantian is not willing to give up some kind of metaphysical realism, namely the belief in a „Ding an sich", a thing in itself – and some philosophers actually insisted that this is superfluous: the German idealists, Friedrich Nietzsche, Arthur Schopenhauer, the phenomenalists etc., see also Cohen 1871, Vaihinger 1911, and Mittelstraß 1977, for instance –, he has to admit that time is a subjective illusion or that there is a dualism between an objective timeless world and a subjective arrow of time; and then a distinction is necessary, which ultimately may lead to the proposal advocated here.)

To get a simplified idea of what is suggested here, take classical thermodynamics and statistical mechanics as an analogy and have a look at figure 2 and 3 (cf. Albrecht 2004).

*Figure 2:*

**Micro- and macrotime in a closed system**

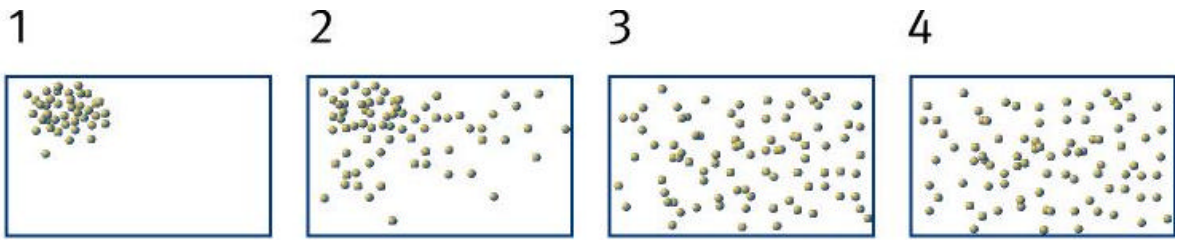

Systems have an arrow of time if they undergo a directed development. This is manifest only on a macroscopic level and is called macroscopic time here (or macrotime for short). It comes along with an increase of entropy which is a measure for the disorder of the system.

For example molecules in a closed box (figure 2) spread from a corner (1) – if they were released there, for instance, from a gas cylinder – in every direction and eventually occupy the whole space (3). Then a state of equilibrium is reached which has no directed development anymore and thus no macrotime. Coarse-grained „snapshots" of the whole system or sufficiently large parts of it show no difference (3 and 4). On a fine-grained level there are still changes (3 versus 4). Thus, a microtime always remains. Due to accidental, sufficiently large fluctuations – which happen statistically even in a state of equilibrium if there is enough microtime available – local structures can arise (from 3 to 1) and a macrotime temporarily comes into being again.

*Figure 3:*

**<u>Micro- and macrotime in an open system</u>**

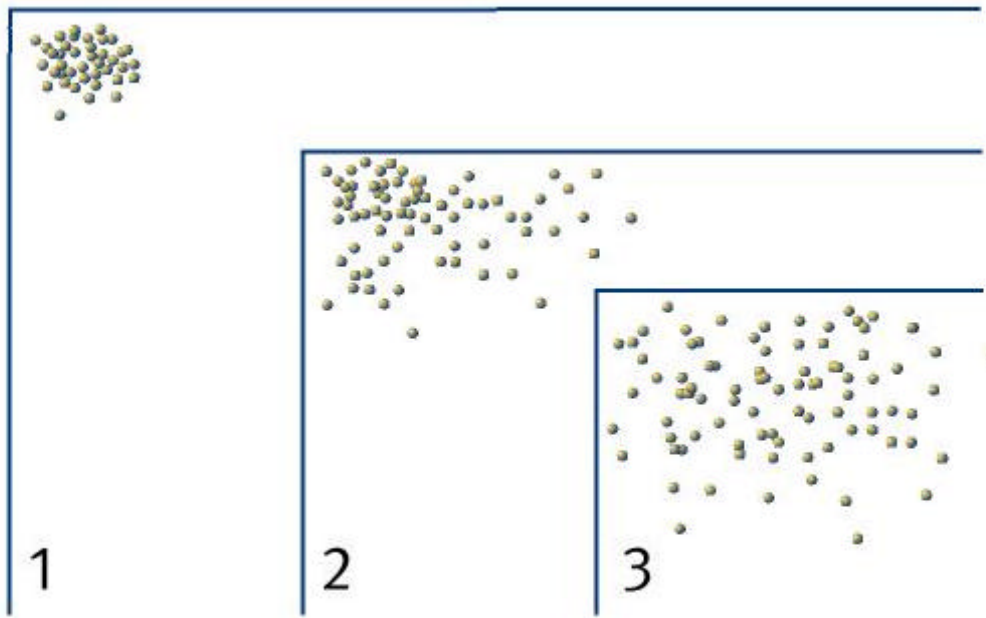

If the system is not closed but open (figure 3), a state of equilibrium does not necessarily develop and macrotime does not vanish. For instance, in the universe this is the case because space expands. Whether there were specific, improbable initial conditions at the big bang (1) or whether order and a directed development could have come out of quite different initial configurations is controversial. Possibly the whole universe is an accidental fluctuation in a macrotimeless quantum vacuum.

In conclusion, and contrary to Kant's thoughts: There are reasons to believe that it is possible, at least conceptually, that time has both a beginning – in the macroscopic sense with an arrow – and is eternal – in the microscopic notion of a steady state with statistical fluctuations.

Is there also some physical support for this proposal?

Surprisingly, quantum cosmology offers a possibility that the arrow has a beginning and that it nevertheless emerged out of an eternal state without any macroscopic time-direction. (Note that there are some parallels to a theistic conception of the creation of the world here, e.g. in the Augustinian tradition which claims that time together with the universe emerged out of a time-less God; but such a cosmological argument is quite controversial, especially in a modern form, cf. Craig & Smith 1993, and of course beyond the scope of this paper.) So this possible overcoming of the first antinomy is not only a philosophical conceivability but is already motivated by modern physics. At least some scenarios of quantum cosmology, quantum geometry/loop quantum gravity, and string cosmology can be interpreted as examples for such a local beginning of our macroscopic time out of a state with microscopic time, but with an eternal, global macroscopic timelessness (further remarks about this models below).

To put it in a more general, but abstract framework and get a sketchy illustration, consider figure 4.

*Figure 4:*

**<u>Something out of almost nothing</u>**

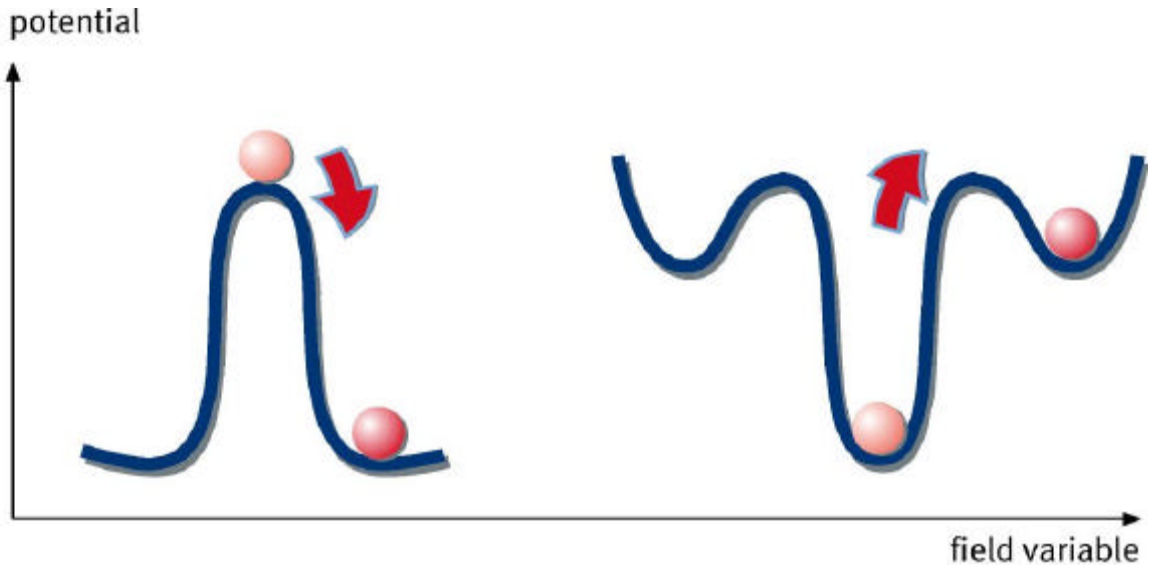

Physical dynamics can be described using „potential landscapes" of fields. For simplicity, here only the variable potential (or energy density) of a single field is shown. To illustrate the dynamics, one can imagine a ball moving along the potential landscape. Depressions stand for states which are stable, at least temporarily. Due to quantum effects, the ball can „jump over" or „tunnel through" the hills. The deepest depression represents the ground state.

In the common theories the state of the universe – the product of all its matter and energy fields, roughly speaking – evolves out of a metastable „false vacuum" (figure 4, left: top) into a „true vacuum" (bottom) which has a state of lower energy (potential). There might exist many (perhaps even infinitely many) true vacua which would correspond to universes with different constants or laws of nature (imagine the graph in figure 4 not as a line but as a 3D shaped object lika a sombrero, where every spot at its brim corresponds to a different universe with a different set of physical laws and constants). The initial condition of this scenario, however, remains unexplainable.

It is more plausible to start with a ground state which is the minimum of what physically can exist (figure 4, right). According to this view an absolute nothingness is impossible. There is something rather than nothing because something cannot come out of absolutely nothing, and something does obviously exist. Thus, something can only change, and this change might be described with physical laws. Hence, the ground state is almost „nothing", but can become thoroughly „something" (more). (Therefore, it is only marginally, but qualitatively wrong that there is something rather than nothing because nothing is instable – a difference which makes everything in the world and, in fact, the whole world.) Possibly, our universe – and, independent from this, many others, probably most of them having different physical properties – arose from such a phase transition out of a quasi atemporal quantum vacuum (and, perhaps, got disconnected completely). Tunneling back might be prevented by the exponential expansion of this brand new space. Because of this cosmic inflation the universe not only became gigantic but simultaneously the potential hill broadened enormously and got (almost) impassable. This preserves the universe from relapsing into its non-existence. On the other hand, if there is no physical mechanism to prevent the tunneling-back or makes it at least very improbable, respectively, there is still another option: If infinitely many universes originated, some of them could be long-lived only for statistical reasons. But this possibility is less predictive and therefore an inferior kind of explanation for not tunneling back.

Another crucial question remains even if universes could come into being out of fluctuations of (or in) a primitive substrate, i.e. some patterns of superposition of fields with local overdensities of energy: Is spacetime part of this primordial stuff or is it also a product of it? Or, more specifically: Does such a primordial quantum vacuum have a semi-classical spacetime structure or is it made up of more fundamental entities?

Both alternatives have already been investigated in some respect. Unique-universe accounts, especially the modified Eddington models – the soft bang/emergent universe mentioned below – presuppose some kind of semi-classical spacetime. The same is true for some multiverse accounts describing our universe (and others) as a fluctuation or collapse/tunnel event – e.g. the models of Edward Tryon (1973), Robert Brout et al. (1978 & 1979), Alexei A. Starobinsky (1980), David Atkatz and Heinz Pagels (1982), John Richard Gott III (1982), and Mark Israelit (2002) –, where Minkowski space, a tiny closed, finite space or the infinite de Sitter space is assumed. The same goes for string theory inspired models like the pre-big bang account of Gabriele Veneziano and Maurizio Gasperini (Gasperini & Veneziano 2003, Vaas 2003a), because string and M-theory is still formulated in a background-dependent way, i.e. requires the existence of a semi-classical spacetime (for some speculative ideas in string theory to go beyond this, i.e. treating spacetime as emergent, see, e.g., Helling 2000, Nishimura 2003). A different approach is the assumption of „building-blocks" of spacetime, a kind of pregeometry (Patton & Wheeler 1975), cf. also the twistor approach of Roger Penrose (cf. Penrose & Rindler 1984 & 1986, Huggett 1998, for a short introduction see Hawking & Penrose 1996, ch. 6), and the cellular automata approach of Stephen Wolfram (2002). The most elaborated accounts in this line of reasoning are quantum geometry (loop quantum gravity) and its relatives (Ashtekar 2002, Ashtekar & Lewandowski 2004, Rovelli 2004, see also Vaas 2003d, and for comparisons with string theory Smolin 2003, Kiefer 2004, Vaas 2004a). Here, „atoms of space and time" are underlying everything.

Though the question whether semiclassical spacetime is fundamental or not is crucial, an answer might be nevertheless neutral with respect of the micro-/macrotime distinction. In both kinds of quantum vacuum accounts the macroscopic time scale is not present. And the microscopic time scale in some respect has to be there, because fluctuations represent change (or are manifestations of change). This change, reversible and relationally conceived, does not occur „within" microtime but constitutes it. Out of a total stasis (even on the microscopic level) nothing new and different can emerge, because an uncertainty principle – fundamental for all quantum fluctuations – would not be realized. In an almost, but not completely static quantum vacuum however, macroscopically nothing changes either, but there are microscopic fluctuations.

Surely a background-independent approach is more promising philosophically, for it is simpler, more fundamental and takes the lesson of general relativity more serious. One can apply the concept of microtime to such a „spacetime dust" nevertheless. A good illustration (or candidate) is the spin foam approach in loop quantum gravity (cf. Baez 2000, Oriti 2003, Perez 2003), but in principle spin networks would also suffice (cf. Rovelli 2004).

To summarize: The pseudo-beginning of our universe (and probably infinitely many others) is a viable alternative both to initial and past-eternal cosmologies and philosophically very significant. Note that this kind of solution bears some resemblance to a possibility of avoiding the spatial part of Kant's first antinomy, i.e. his claimed proof of both an infinite space without limits and a finite, limited space: The theory of general relativity describes what was considered logically inconceivable before, namely that there could be universes with finite, but unlimited space (Einstein 1917), i.e. this part of the antinomy also makes the error of the excluded third option. This offers a middle course between the Scylla of a mysterious, secularized *creatio ex nihilo*, and the Charybdis of an equally inexplicable eternity of the world.

In this context it is also possible to defuse some explanatory problems of the origin of „something" (or „everything") out of „nothing" (cf. Nozick 1981 & 2001, ch. 3, Parfit 1998) as well as a – merely assumable, but never provable – eternal cosmos or even an infinitely often recurring universe. But that does not offer a final explanation or a sufficient reason, and it cannot eliminate the ultimate contingency of the world.

# 5. Discussion

Before coming back to conceptual questions and cosmological classifications, a possible objection against what was proposed here shall briefly be discussed. If our universe has a beginning within the multiverse, one could argue that it was only stated here that *certain parts of the world* had a beginning, but not *the world as a whole*, i.e. the sum of all its parts. Thus, the question would repeat itself, although in a larger context. And, indeed, this is already an issue of contemporary cosmology.

For example Alan Guth, Alvin Borde and Alexander Vilenkin (2003) argued that within the framework of a future-eternal inflationary multiverse, as well as some more speculative string-cosmologies, *all* worldlines are geodesically incomplete and, thus, the multiverse has to have a beginning. Unfortunately, if future-eternal inflation is true, all „hypotheses about the ultimate beginning of the universe would become totally divorced from any observable consequences. Since our own pocket universe would be equally likely to lie anywhere on the infinite tree of universes produced by eternal inflation, we would expect to find ourselves arbitrarily far from the beginning. The infinite inflating network would presumably approach some kind of steady state, losing all memory of how it started […] Thus, there would be no way of relating the properties of the ultimate origin to anything that we might observe in today's universe" (Guth 2001, p. 78).

On the other hand, Andrei Linde and – at least for the sake of argument – Anthony Aguirre and Steven Gratton argued that the multiverse could be past-eternal, because either all single world lines might have to start somewhere, but not the whole bundle of them (Linde 2004), or there could even exist some (albeit strange) space-times with single past-eternal world lines (Aguirre & Gratton 2002).

This issue is not settled, and even in those scenarios a global arrow of time may not necessarily exist. However, there are other frameworks possible – and they have even already been developed to some extent –, where a future-eternal inflationary multiverse is both not past-eternal and beginningless but arise from some primordial vacuum which is macroscopically time-less. Thus, again, the beginning of some classical space-times is not equivalent with the beginning of everything.

We can even imagine that there is no multiverse, but that the whole (perhaps finite) universe – our universe – once was in a steady state without any macroscopic arrows of time but, due to a statistical fluctuation above a certain threshold value, started to expand (Rebhan 2000 and, independently, Barrow et al. 2003, Ellis & Maartens 2004, Ellis et al. 2004) – or to contract, bounce and expand – as a whole and acquired an arrow of time. In such a case the above-mentioned reply, which was based on the spatial distinction of a beginning of some parts of the world and the eternity of the world as a whole, would collapse.

Nevertheless it is necessary to distinguish between the different notions and extensions of the term „universe". In the simplest case, Kant's antinomy might be based on an ambiguity of the term „world" (i.e. the difference between „universe" and „multiverse"), but it does not need to; and it was not assumed here that it necessarily does.

The temporal part of Kant's first antinomy was purely about the question whether the macroscopic arrow of time is past-eternal or not. And if it is not past-eternal this does not mean that time and hence the world has an absolute beginning in every respect – it is still possible that there was or is a world with some underlying microscopic time. (By the way, one can also imagine that, even if our arrow of time is past- and/or future-eternal, there might exist „timeless islands" someday: for instance isolated black holes if they would not ultimately radiate away due to quantum effects, or empty static universes if they could split off of our space-time.)

Of course it is possible that firstly a natural *principle of plentitude* is realized and different multiverses (sets of universes) exist totally independent from each other, and secondly that some of them are truly past-eternal while others have an absolute beginning and others have only local starting points of local arrows of time as it was suggested here. If so, we might not be able to tell in what kind we live in. And this would be irrelevant in the end, because then every possible world is actual and probably exists infinitely often. But we do not know whether such an extreme principle of plentitude does apply or if cosmology is ultimately just and only a matter of pure logical consistency, allowing us finally to calculate the complete architecture of the world by armchair-reasoning. In any case, one goal of this paper is a more modest one: to show that there is a promising third option besides Kant's dichotomy and antinomy, but neither to prove that such a possibility is the only consistent one nor that, as a matter of fact, our universe arose from such a time before time.

## 6. Some problems of conceptual classifications in cosmology

Before concluding with some more general remarks, the remainder of this paper shall address some open issues of table 3 above, i.e. the classification of different cosmological models. Here the distinction between microscopic and macroscopic time might also be useful.

• The classical Phoenix universe (Lemaître 1933, Tolman 1934), a future-eternal big bang-big crunch sequence, is globally (!) characterized here as an initial cosmology. (Of course locally, i.e. viewed from within, it has a singular beginning and an end, for one cannot pass into the next cycle; so one could see the model either as one oscillating universe or as a sequence of different spatially closed universes both connected and separated by a singularity.) The reason for this is that due to entropy considerations the photons-to-baryons-ratio increases from each cycle to the next and therefore a later cycle lasts longer than an earlier one and the maximum volume is increasing. If one does not assume an infinite continuum of smaller and smaller cycles in the past, a very brief first big bang-big crunch episode must have started the whole series. (This might also be the case for Lee Smolin's cosmic Darwinism, a future-eternal realm of universes branching off their parent universe out of black holes or a big crunch, see Smolin 1992 & 1997 & 2004b, cf. Vaas 1998 & 2003e, Susskind 2004b). The Phoenix scenario (which does not address the dark energy problem however) was seen as an alternative to cosmic inflation (Durrer & Laukenmann 1996).

• Note that the string theory-inspired cyclic universe model of Paul Steinhardt and Neil Turok et al. (Khoury, J. et al 2004, Steinhardt & Turok 2002abc, Turok & Steinhardt 2002, 2004ab, see also Vaas 2001c & 2002b), which also is a viable alternative to inflation and can be seen from an effective 4D field theory perspective as an oscillating universe too, does not have such an entropy problem, because entropy is inflated away during the accelerated expansion of the brane – therefore global entropy does increase, but entropy density doesn't. However it is nevertheless not clear whether this cyclic universe has an infinite past (for counter arguments see Borde, Guth & Vilenkin 2003, Albrecht 2004, cf. also Steinhardt & Turok 2004a, for a general critique e.g. Linde 2003, for a loop quantum cosmological approach Bojowald, Maartens & Singh 2004). In any case, in contrast to the Phoenix universe, the cyclic universe is really cyclic, i.e. every epoch is, at least globally, a recurrence of the earlier ones. By the way, even if the cyclic sequence is not past-eternal, this model should be viewed as an eternal cosmology nevertheless. This is because from the 5D perspective the spatially infinite four-dimensional branes, which are separated by a five-dimensional bulk and collide (or pass through each other) periodically, are conceived as past- and future-eternal themselves – even if their cyclic behavior is not. (However, it is possible in principle that brane worlds do have an absolute beginning, i.e. come out of a Hawking instanton (cf. Hawking, Hertog & Reall 2000, see also Hawking 2001, ch. 7 & 2003), but there is no model yet which combines such an initial brane cosmology with a future-eternal cyclic brane cosmology.)

• Globally, the Planckian cosmic egg model also can be classified as an oscillating universe. It starts with a nonsingular, closed universe filled with an ultra-dense (Planck density) and extremely cold ($10^{-180}$ Kelvin) pre-matter with an extreme tension leading to inflation (Israelit & Rosen 1989, extensions with a scalar-field are given by Starkovich & Cooperstock 1992 and Bayin, Cooperstock & Faraoni 1994). This unique universe expands and contracts to a new egg which inflates again ad infinitum. However, the Planckian egg is just an axiomatic assumption. More recent work try to explain it as the result of a primary pre-Planckian geometric entity without matter, a quasi-static closed space with some fields (Israelit 2002). Under the even more speculative assumption of an Integrable Weyl-Dirac theory this would lead to the Planckian egg and can be interpreted as a pseudo-beginning universe.

• While the above-mentioned cycles might be more or less identical, they are not identified, that is the same, but a linear sequence in global time (at least from the theoretical or God's eye view perspective). However, rotating Gödel universes could be interpreted differently, for they contain globally closed timelike loops. In this case, there might be only one cycle which is repeating itself infinitely often – i.e. time is circular, not linear. However, not all worldlines are temporally closed (indeed, most are not), and in the original model geodesics are not temporally closed at all (only sufficiently large accelerated movements lead into one's own past). But there are more complex Gödel world models (e.g. with electromagnetic fields, cf. Earman 1995, p. 168), which have such closed geodesics. On the other hand, expanding and inhomogeneous Gödel universes are also possible, which do not have identifiable cycles, i.e. only the same repeating itself without end.

• A very different cyclic kind of model assumes a time reversal when space stops to expand and starts to contract again (Kiefer & Zeh 1995, cf. Vaas 2002c). Thus, in principle time could „run" endlessly forward and backward between the big bang and the big crunch, but this arrow switch cannot be observed (the psychological arrow of time does also switch), so both end points could literally be identified as the same – humorously called a „big brunch".

• Another time-loop model assumes a cyclic time only at the beginning of the universe or multiverse (Gott & Li 1998, cf. Vaas 2004d). Here the universe could be its own mother, i.e. it is self-created. Time, as it were, circles in such a tiny, spatially closed adopted Rindler vacuum (a low entropy state due to zero Kelvin temperature), but as soon as it „escapes", so to speak, passing a Cauchy horizon, irreversibility starts, i.e. an arrow of time. Thus, such a self-creating universe might also be seen as a pseudo-beginning universe but of a different kind than the other models mentioned in table 3.

• It could be argued that the classical Eddington universe solutions are so close to the soft bang/emergent universe models (for they are in fact Eddington solutions combined with cosmic inflation) that they should not be classified into different categories. Historically, Eddington universes – like Einstein's static universe – were taken as past-eternal, i.e. with an infinite past. However, with regard to the distinction between microscopic and macroscopic time one could argue that within the static realm there is/was no macroscopic time, i.e. no arrow of time. From that line of reasoning one might conclude that Eddington universes do also have a pseudo-beginning. Or that the static Einstein universe has, in a sense, no macroscopic time at all, i.e. it would be a timeless universe (which is not to be confused with a block universe view, for those interpretations could be applied to all initial and eternal cosmologies). However, one should not confuse mathematical time (which is a kind of absolute time within the abstract description of the world model) with physical time – whatever this exactly means. As Eckard Rebhan (2000, p. 9) wrote: „In physics time is a parameter that is used for ordering changes of states. However, when there are no changes then this order parameter loses its sense. In a very slowly changing situation it may therefore become more useful to consider the changes themselves as the order parameter that represents time instead of using an order parameter ordering no changes." In this sense the macroscopic lifetime of the universe considered in soft bang/emergent universe models is not greater than that of classical big bang models. (One might still argue that the Eddington instability developed out of an infinite past, thus there is still a cosmological arrow of time and, hence, soft bang/emergent universe models do not have a pseudo-beginning. But this is taken as physically misleading here and might just be a mathematical artifact.) – So the classifications depend on details which are still ignored by the very simplified models. If isotropy and homogeneity were perfect (i.e. not just a sufficiently coarse-grained simplification as a boundary condition to solve the equations) and all matter was literally taken as a cosmic fluid, for instance, such universes could not represent our own with its complex structure (galaxies etc.) on a more fine-grained level of description. But the Einstein and Eddington solutions – just like the steady state model – were not meant as universes without such structures. The soft bang/emergent universe models, however, do assume such a very simple state before an instability occurs, because ordinary matter was not there. Matter was created only after the inflationary epoch (due to inflaton decay, reheating etc.). That's why those models were classified as pseudo-beginning universes here.

• A related problem refers to the pre-big bang model (Veneziano & Gasperini 2003). Here the string vacuum – where a local collapse in the Einstein frame (which corresponds to a dilaton-driven inflation in the string frame) before the big bang occurs – is quite simple, homogeneous, almost empty and does not have an overall arrow of time. But, mathematically, the origin of the pre-big bang – or, to be more precise, any pre-big bang, for the model does also imply a multiverse scenario – traces back to a maximally simple, static condition only in the infinite past (principle of asymptotic past triviality). But this can also be interpreted just as a local predecessor of a big bang and not a feature characterizing the infinite string vacuum as a whole. „The universe becomes more and more cold, empty, flat and non-interacting as time becomes more and more negative, until it reaches complete emptiness and triviality in its asymptotic past. The universe would thus obey a principle of Asymptotic Past Triviality, emerging from the simplest possible kind of initial states" (Veneziano, p. 19). „It seems indeed physically (and philosophically) satisfactory to identify the beginning with simplicity. However, simplicity should not be confused with complete triviality: a rigorously empty and flat Universe, besides being uninteresting, is also very special, i.e. non-generic. By contrast, asymptotically trivial Universes, though initially simple, are also generic in a precise mathematical sense" (Gasperini & Veneziano 2003, online p. 53).

• The big bounce model of Hans-Joachim Blome and Wolfgang Priester (1991, cf. Vaas 1994) might also be classified as a pseudo-beginning universe if the state before the bounce is interpreted as a quantum vacuum without a macroscopic time. However, this model still has a cosmological arrow of time, namely the shrinking scale-factor, i.e. a contraction of space from the infinite past. But one might reply that it could not be observed within such a pre-bounce universe for this stage is empty (matter is only created after the brief epoch of inflation occurring after the bounce). Until a more detailed description of the pre-bounce state is given, probably depending on a theory of quantum gravity, it has to be admitted that the classification is somewhat arbitrary.

• On the other hand, the singularity-free loop quantum cosmology model developed by Martin Bojowald (2004 and Bojowald & Morales-Tecotl 2004 for reviews, cf. Ashtekar 2002, Vaas 2004b), classified here as a pseudo-beginning cosmology, also undergoes a kind of bounce. However, the fabric of space and time literally dissolves around the bounce. It is an open issue, whether there was a semi-classical spacetime „before" it. More likely, there was a strange quantum state, describable only in an abstract manner with spin networks or spin foams in superposition. (Of course all these questions depend on the notorious problem of time in quantum gravity, which is not solved, c.f., e.g., Smolin 2000.) „The question of whether the universe had a beginning at a finite time is now ‚transcended'. At first, the answer seems to be ‚no' in the sense that the quantum evolution does not stop at the big-bang. However, since space-time geometry ‚dissolves' near the big-bang, there is no longer a notion of time, or of ‚before' or ‚after' in the familiar sense. Therefore, strictly, the question is no longer meaningful. The paradigm has changed and meaningful questions must now be phrased differently, without using notions tied to classical space-times" (Ashtekar, Bojowald & Lewandowski 2003).

To summarize, the classification of table 3 should not be viewed as the ultimate truth, but as a proposal which might be subject to change both due to new interpretations in the future and more detailed and realistic modifications and additions of the models in question.

## 7. Outlook

It seems unlikely that philosophical considerations alone can answer the question whether there was a beginning of the universe or not, and in what sense. It is also premature, however, to ignore such questions, e.g. for Kantian reasons. On the other hand, it is not to be expected that some unambiguous empirical results (e.g. from the gravitational wave background, dark matter relics, or some traces in the cosmic background radiation) will ever solve these questions. Empirical research might at least constrain cosmological theories which could and should be based on current and/or future and more advanced fundamental theories of forces, particles, space and time, e.g. M-theory or quantum geometry (cf. Smolin 2000 & 2003, Vaas 2004a). It is an open question whether the dream of such a „final theory" (Weinberg 1993) or „Theory of Everything" (Barrow 1990) will ultimately explain the origin of our universe (or even the whole multiverse) and address the finiteness or infinity of space and time – or even reduce space-time to something more fundamental. But extrapolating from the scenario

– 20 –of eternal inflation (Guth 2000ab & 2001 & 2002, Vilenkin 2000), and contemporary approaches to quantum gravity, it seems almost inevitable that the origin of our universe was *not* a unique event, and that other universes also exist. This not only has important implications for observational (or anthropic) selection effects (Barrow and Tipler 1986, Kanitscheider 2001, Vaas 2000 & 2004e) and our place in nature (Knobe, Olum & Vilenkin 2003, Vaas 2001a), but also for the question whether the whole cosmos or multiverse is past-eternal or not (Borde, Guth & Vilenkin 2003). If the proposal of this paper is correct, both options could be true in some way.

# References

Abbott, L. F., Pi, S.-Y. (eds.) 1986: Inflationary Cosmology. Singapore: World Scientific.
Albrecht, A. 2004: Cosmic Inflation and the Arrow of Time. In: Barrow, J. D., Davies, P. C. W., Harper, C. L. (eds.): Science and Ultimate Reality. Cambridge: Cambridge University Press, http://arXiv.org/abs/astro-ph/0210527
Ashtekar, A. 2002: Quantum Geometry In Action: Big Bang and Black Holes, http://arXiv.org/abs /math-ph/0202008
Ashtekar, A., Bojowald, M., Lewandowski, J. 2003: Mathematical structure of loop quantum cosmology. Advances in Theoretical and Mathematical Physics 7, 233-268, http://arxiv.org/abs/gr-qc/0304074
Ashtekar, A., Lewandowski, J. 2004: Background Independent Quantum Gravity: A Status Report, http://arxiv.org/abs/gr-qc/0404018
Atkatz, D., Pagels, H. R. 1982: Origin of the Universe as a Quantum Tunneling Event. Phys. Rev. D25, 2065-2072.
Atkins, P. W. 1981: The Creation. Oxford: W. H. Freeman.
Baez, J. C. 2000: An Introduction to Spin Foam Models of Quantum Gravity and BF Theory. Lect. Notes Phys. 543, 25-94, http://arxiv.org/abs/gr-qc/9905087
Barbour, I. 2000: The End of Time. Oxford, New York: Oxford University Press.
Barrett, J. A. 1999: The Quantum Mechanics of Minds and Worlds. Oxford: Oxford University Press.
Barrow, J. D. 1990: Theories of Everything. Oxford: Oxford University Press.
Barrow, J. D., Tipler, F. J. 1986: The Anthropic Cosmological Principle. Oxford: Oxford University Press.
Barrow, J. D., Ellis, G., Maartens, R., Tsagas, C. 2003: On the Stability of the Einstein Static Universe. Class. Quant. Grav. 20, L155-L164, http://arXiv.org/abs /gr-qc/0302094
Bartels, A. 1996: Grundprobleme der modernen Naturphilosophie. Paderborn u. a.: Schöningh.
Bayin, S. S., Cooperstock, F. I., Faraoni, V. 1994: A singularity-free cosmological model with a conformally coupled scalar field. Astrophys. J. 428, 439-446.
Blau, S. K., Guth, A. H. 1987: Inflationary cosmology. In: Hawking, S. W., Israel, W. (eds.): Three hundred years of gravitation. Cambridge: Cambridge University Press 1989, 524-603.
Blome, H. J., Priester, W. 1991: Big Bounce in the Very Early Universe. Astron. Astrophys. 250, 43-49.
Bojowald, M. 2004: Loop Quantum Cosmology: Recent Progress. Plenary talk at ICGC 04, Cochin, India. http://arxiv.org/abs/gr-qc/0402053
Bojowald, M., Morales-Tecotl, H. A. 2004: Cosmological applications of loop quantum gravity. Lecture Notes in Physics 646, 421-462, http://arxiv.org/abs/gr-qc/0306008
Bojowald, M., Maartens, R., Singh, P. 2004: Loop Quantum Gravity and the Cyclic Universe, http://arxiv.org/abs/hep-th/0407115
Borde, A., Guth, A. H., Vilenkin, A. 2003: Inflationary spacetimes are not past-complete. Phys. Rev. Lett. 90, 151301, http://arXiv.org/abs/gr-qc/0110012
Brout, R., Englert, F., Gunzig, E. 1978: The Creation of the Universe as a Quantum Phenomenon. Annals of Physics 115, 78-106.
Brout, R., Englert, F., Spindel, P. 1979: Cosmological Origin of the Grand-Unification Mass Scale. Phys. Rev. Lett. 43, 417-420.
Carr, B. J. 2005: Universe or Multiverse. Cambridge: Cambridge University Press.
Cohen, H. 1871: Kants Theorie der Erfahrung. Berlin: Dümmler.
Craig, W. L., Smith, Q. 1993: Theism, atheism, and big bang cosmology. Oxford: Clarendon Press.
Davies, P. C. W. 2004: Multiverse cosmological models. Mod. Phys. Lett. A19, 727-744, http://arxiv.org/abs/astro-ph/0403047

Deutsch, D. 1997: The Fabric of Reality. London, New York: Allen Lane.
Deutsch, D. 2001: The Structure of the Multiverse, http://arxiv.org/abs/quant-ph/0104033
DeWitt, B. S., Graham, N. (eds.) 1973: The Many-Worlds Interpretation of Quantum Mechanics. Princeton: Princeton University Press.
Douglas, M. R. 2003: The statistics of string/M theory vacua, http://arxiv.org/abs/hep-th/0303194
Douglas, M. R. 2004: Statistical analysis of the supersymmetry breaking scale, http://arxiv.org/abs/hep-th/0405279
Durrer, R., Laukenmann, J. 1996: The Oscillating Universe: an Alternative to Inflation. Class. Quant. Grav. 13, 1069-1088. http://arxiv.org/abs/gr-qc/9510041
Earman, J. 1995: Bangs, Crunches, Whimpers, and Shrieks. New York, Oxford: Oxford University Press.
Eddington, A. S. 1930: On the Instability of Einstein's Spherical World. M. N. R. A. S. 19, 668-678.
Einstein, A. 1917: Kosmologische Betrachtungen zur allgemeinen Relativitätstheorie. Königlich Preußische Akademie der Wissenschaften (Berlin). Sitzungsberichte, 142-152. Reprinted in: Knox, A. J., Klein, M. J., Schulmann, R. (eds.): The Collected Papers of Albert Einstein. Princeton: Princeton University Press 1996, Vol. 6, 541-552.
Ellis, G. F. R., Kirchner, U., Stoeger, W. R. 2004: Multiverses and physical cosmology. M. N. R. A. S. 347, 921-936, http://arxiv.org/abs/astro-ph/0305292
Ellis, G., Maartens, R. 2004: The Emergent Universe: inflationary cosmology with no singularity. Class. Quant. Grav. 21, 223-232, http://arXiv.org/abs /gr-qc/0211082
Ellis, G., Maartens, R., Murugan, J., Tsagas, C. 2004: The Emergent Universe: An Explicit Construction. Class. Quant. Grav. 21, 233-250, http://arXiv.org/abs /gr-qc/0307112
Fakir, R. 2000: General Relativistic Cosmology With No Beginning of Time. Astrophys. J. 537, 533-536, http://arxiv.org/abs/gr-qc/9810054
Falkenburg, B. 2000: Kants Kosmologie. Frankfurt am Main: Klostermann.
Friedmann, A. 1922: Über die Krümmung des Raumes. Zeitschrift für Physik 10, 377-386.
Friedmann, A. 1924: Über die Möglicheit einer Welt mit konstanter negativer Krümmung des Raumes. Zeitschrift für Physik 21, 326-332.
Gasperini, M., Veneziano, G. 2003: The Pre-Big Bang Scenario in String Cosmology. Phys. Reports 373, 1-212, http://arXiv.org/abs/hep-th/0207130
Gell-Mann, M., Hartle, J. 1990: Quantum Mechanics in the Light of Quantum Cosmology. In: Zurek, W. H. (ed.): Compexity, Entropy, and the Physics of Information. Redwood City: Addison-Wesley, 425-458.
Gell-Mann, M., Hartle, J. 1993: Time Symmetry and Asymmetry in Quantum Mechanics and Quantum Cosmology. In: Halliwell, J., Perez-Mercader, J., Zurek, W. (eds.): Physical Origins of Time Asymmetry. Cambridge: Cambridge University Press, 311-345, http://arXiv.org/abs/gr-qc/9304023
Gödel, K. 1949: An Example of a New Type of Cosmological Solutions of Einstein's Field Equations. Rev. Mod. Phys. 21, 447-450.
Gödel, K. 1952: Rotating Universes in General Relativity Theory. Proceedings of the International Congress of Mathematicians. Providence: American Mathematical Society, vol. 1, 175-181.
Gott III, J. R. 1982: Creation of Open Universes from de Sitter Space. Nature 295, 304-307.
Gott III, J. R., Li, L.-X. 1998: Can the Universe Create Itself? Phys. Rev. D58, 023501, http://arxiv.org/abs/astro-ph/9712344
Gouts, A. K. 2003: The theory of Multiverse, multiplicity of physical objects and physical constants. Grav. Cosmol. 9, 33-36, http://arxiv.org/abs/gr-qc/0210072
Guth, A. H. 1997: The inflationary universe. Reading: Perseus.
Guth, A. H. 2000a: Inflation and Eternal Inflation. Phys. Reports 333, 555-574, http://arXiv.org/abs/astro-ph/0002156
Guth, A. H. 2000b: Inflationary Models and Connections to Particle Physics, http://arxiv.org/abs/astro-ph/0002188
Guth, A. H. 2001: Eternal Inflation. In: Miller, J. B. (ed.): Cosmic Questions. New York Academy of Sciences: New York, 66-82, http://arXiv.org/abs/astro-ph/0101507
Guth, A. H. 2002: Time Since the Beginning, http://arxiv.org/abs/astro-ph/0301199
Grünbaum, A. 1991: Die Schöpfung als Scheinproblem der physikalischen Kosmologie. In Bohnen, A., Musgrave, A. (Hrsg.): Wege der Vernunft. Tübingen: Mohr, 164-191.
Hartle, J., Hawking, S. W. 1983: The wave function of the universe. Phys. Rev. D28, 2960-2975.
Hawking, S. W. 1988: A Brief History of Time. New York: Bantam.
Hawking, S. W. 2001: The Universe in a Nutshell. New York: Bantam Books.

Hawking, S. W. 2003: Cosmology from the Top Down, http://arxiv.org/abs/astro-ph/0305562

Hawking, S. W., Hertog, T., Reall, H. S. 2000: Brane New World. Phys. Rev. D62, 043501, http://arxiv.org/abs/hep-th/0003052

Hawking, S. W., Penrose, R. 1970: The Singularities of Gravitational Collapse and Cosmology. Proc. Royal. Soc. London A314, 529-548.

Hawking, S. W., Penrose, R. 1996: The Nature of Space and Time. Princeton: Princeton University Press.

Hawking, S. W., Turok, N. 1998: Open Inflation Without False Vacua. Phys. Lett. B425, 25-32, http://arXiv.org/abs/hep-th/9802030

Heimsoeth, H. 1960: Zeitliche Weltunendlichkeit und das Problem des Anfangs. In: Studien zur Philosophiegeschichte. Köln: Kölner Universitäts-Verlag 1961, 269-292.

Helling, R. 2000: D-Geometry, http://www.damtp.cam.ac.uk/user/rch47/schloessmannposter.ps.gz

Hilbert, D. 1925: Über das Unendliche. In: Math. Annalen 95, 161-190. Engl. transl.: On the Infinite. In: Benacerraf, P., Putnam, H. (eds.): Philosophy of Mathematics. Englewood Cliff: Prentice Hall 1964, 134-151.

Hoyle, F., Burbidge, G., Narlikar, J. V. 2000: A Different Approach to Cosmology. Cambridge: Cambridge University Press.

Huggett, S. A., et al. (eds.) 1998: The Geometric Universe. Oxford: Oxford University Press.

Israelit, M. 2002: Primary Matter Creation in a Weyl-Dirac Cosmological Model. Foundations of Physics 32, 295-321.

Israelit, M., Rosen, N. 1989: A singularity-free cosmological model in general relativity. Astrophys. J. 342, 627-634.

Kanitscheider, B. 1991: Kosmologie. Stuttgart: Reclam, 2. erw. Aufl.

Kanitscheider, B. 2001: Die Feinabstimmung des Universums. In: Meixner, U. (Hrsg.): Metaphysik im postmetaphysischen Zeitalter/Metaphysics in the Post-Metaphysical Age. Wien: öbv & hpt, 207-217.

Kant, I. 1781/1787: Kritik der reinen Vernunft. Frankfurt am Main: Suhrkamp 1990; engl:. Critique of Pure Reason. E.g. translated by N. Kemp Smith, http://www.arts.cuhk.edu.hk/Philosophy/Kant/cpr/

Khoury, J. et al. 2001: The Ekpyrotic Universe: Colliding Branes and the Origin of the Hot Big Bang. Phys. Rev. D64, 123522, http://arxiv.org/abs/hep-th/0103239

Khoury, J., Steinhardt, P. J., Turok, N. 2003: Inflation versus Cyclic Predictions for Spectral Tilt. Phys. Rev. Lett. 91, 161301, http://arxiv.org/abs/astro-ph/0302012

Khoury, J., Steinhardt, P. J., Turok, N. 2004: Designing Cyclic Universe Models, http://arxiv.org/abs/hep-th/0307132

Kiefer, C. 2004: Quantum Gravity. Oxford: Oxford University Press.

Knobe, J., Olum, K. D., Vilenkin, A. 2003: Philosophical Implications of Inflationary Cosmology, http://arXiv.org/abs/physics/0302071

Kragh, H. 1996: Cosmology and Controversy. Princeton: Princeton University Press.

Lemaître, G. 1927: Un univers homogène de masse constante et de rayon croissant, rendant compte de la vitesse radiale des nébuleuses extragalactiques. Ann. Soc. Sci. Bruxelles A47, 49-59.

Lemaître, G. 1933: L'univers en expansion. Ann. Soc. Sci. Bruxelles A53, 51-85.

Lidsey, J. E. et al. 1997: Reconstructing the inflation potential – an overview. Rev. Mod. Phys. 69, 373-410.

Linde, A. 1983: Chaotic inflation. Phys. Lett. B129, 177-181.

Linde, A. 1994: The Self-Reproducing Inflationary Universe. Scientific American 271 (5), 48-55, http://physics.stanford.edu/linde/1032226.pdf

Linde, A. 2003a: Inflation, Quantum Cosmology and the Anthropic Principle. In: Barrow, J. D., Davies, P. C. W., Harper, C.L. (eds.): Science and Ultimate Reality. Cambridge: Cambridge University Press, http://arxiv.org/abs/hep-th/0211048

Linde, A. 2003b: Inflationary theory versus ekpyrotic/cyclic scenario. In: Gibbons, G. W., Shellard, E. P. S., Rankin,, S. J. (eds.): The Future of Theoretical Physics and Cosmology. Cambridge: Cambridge University Press, 801-838, http://arxiv.org/abs/hep-th/0205259

Linde, A. 2004: Prospects of Inflation, http://arxiv.org/abs/hep-th/040

Malzkorn, W. 1999: Kants Kosmologie-Kritik. Berlin, New York: Walter de Gruyter.

Maudlin, T. 2002: Remarks on the passing of time. Proc. Arist. Soc. CII, 237-252.

Mittelstraß, J. 1977: Ding als Erscheinung und Ding an sich. In: Mittelstraß, J., Riedel, M. (Hrsg.): Vernünftiges Denken. Berlin, New York: de Gruyter.

Nishimura, J. 2003: Lattice Superstring and Noncommutative Geometry, http://arxiv.org/abs/hep-lat/0310019

Nozick, R. 1981: Why is there something rather than nothing? In: Philosophical Explanations. Oxford: Clarendon Press, 115-164, 668-679.

Nozick, R. 2001: Invariances. Cambridge, London: Harvard University Press.
Oriti, D. 2003: Spin Foam Models of Quantum Spacetime, http://arxiv.org/abs/gr-qc/0311066
Parfit, D. 1998: Why anything? Why this? I & II. London Review of Books, Jan. 22 & Feb. 5, 24-27 & 22-25.
Patton, C. M., Wheeler, J. A. 1975: Is physics legislated by cosmogony? In: Isham, C. J., Penrose, R., Sciama, D. W. (eds.): Quantum gravity. Oxford: Clarendon Press, 538-605.
Peiris, H. V. E., et al. 2003: First Year Wilkinson Microwave Anisotropy Probe (WMAP) Observations: Implications for Inflation. Astrophys. J. Suppl. 148, 213-231, http://arxiv.org/abs/astro-ph/0302225
Penrose, R., Rindler, W. 1984 & 1986: Cambridge: Cambridge University Press, 2 vol.
Penrose, R. 1989: The Emperor's New Mind. Oxford: Oxford University Press.
Penrose, R. 2004: The Road to Reality. London: Jonathan Cape.
Perez, A. 2003: Spin Foam Models for Quantum Gravity. Class. Quant. Grav. 20, R43-R104, http://arxiv.org/abs/gr-qc/0301113
Price, H. 1996: Time's Arrow and Archimedes' Point. New York, Oxford: Oxford University Press.
Prigogine, I. 1979: Vom Sein zum Werden. München, Zürich: Piper.
Rebhan, E. 2000: „Soft bang" instead of „big bang": model of an inflationary universe without singularities and with eternal physical past time. Astron. Astrophys. 353, 1-9.
Rees, M. 2001: Our Cosmic Habitat. Princeton: Princeton University Press.
Rovelli, C. 2004: Quantum Gravity. Cambridge: Cambridge University Press, http://www.cpt.univ-mrs.fr/~rovelli/book.pdf
Schmucker, J. 1990: Das Weltproblem in Kants Kritik der reinen Vernunft. Bonn: Bouvier.
Schulman, L. S. 1997: Time's arrows and quantum measurement. Cambridge: Cambridge University Press.
Sitter, W. de 1917: On the Relativity of Intertia. Koninglijke Nederlandse Akademie van Wetenschappen Amsterdam. Proceedings of the Section of Science 19, 1217-1225.
Smith, Q. 1985: Kant and the Beginning of the World. The New Scholasticism 59, 339-346, http://www.qsmithwmu.com/kant_and_the_beginning_of_the_world.htm
Smith, Q. 2002: Time Was Created by a Timeless Point. In: Ganssle, G. E, Woodruff, D. M. (eds.): God and Time. Oxford, New York: Oxford University Press, http://www.qsmithwmu.com/time_began_with_a_timeless_point.htm
Smolin, L. 1992: Did the universe evolve? Class. Quant. Grav. 9, 173-191.
Smolin, L. 1997: The Life of the Cosmos. New York, Oxford: Oxford University Press.
Smolin, L. 2000: The present moment in quantum cosmology: challenges to the arguments for the elimination of time. In: Durie, R. (ed.): Time and the Instant. Manchester: Clinamen Press, http://arxiv.org/abs/gr-qc/0104097
Smolin, L. 2000: Three Roads To Quantum Gravity. London: Phoenix 2001.
Smolin, L. 2003: How far are we from the quantum theory of gravity?, http://arxiv.org/abs/hep-th/0303185
Smolin, L. 2004a: Atoms of Space and Time. Scientific American 290 (1), 56-65.
Smolin, L. 2004b: Scientific alternatives to the anthropic principle, http://arxiv.org/abs/hep-th/0407213
Spergel, D. N., et al. 2003: First Year Wilkinson Microwave Anisotropy Probe (WMAP) Observations: Determination of Cosmological Parameters. Astrophys. J. Suppl. 148, 175-194, http://arxiv.org/abs/astro-ph/0302209
Spitzer, R. J. 2000: Definitions of Real Time and Ultimate Reality. Ultimate Rality and Meaning 23 (3), 260-276.
Starkovich, S. P., Cooperstock, F. I. 1992: A cosmological field theory. Astrophys. J. 398, 1-11.
Starobinsky, A. A. 1980: A new type of cosmological models without singularity. Phys. Lett. 91B, 99-102.
Steinhardt, P. J., Turok, N. 2002a: A Cyclic Model of the Universe. Science 296, 1436-1439, http://arXiv.org/abs/hep-th/0111030
Steinhardt, P. J., Turok, N. 2002b: Cosmic Evolution in a Cyclic Universe. Phys. Rev. D65, 126003, http://arxiv.org/abs/hep-th/0111098
Steinhardt, P. J., Turok, N. 2002c: Is Vacuum Decay Significant in Ekpyrotic and Cyclic Models?
Phys. Rev. D66, 101302, http://arxiv.org/abs/astro-ph/0112537
Stoeger, W. R., Ellis, G. F. R., Kirchner, U. 2004: Multiverses and Cosmology: Philosophical Issues, http://arxiv.org/abs/astro-ph/0407329
Susskind, L. 2003: The Anthropic Landscape of String Theory, http://arxiv.org/abs/hep-th/0302219
Susskind, L. 2004a: Supersymmetry Breaking in the Anthropic Landscape, http://arxiv.org/abs/hep-th/0405189
Susskind, L. 2004b: Cosmic Natural Selection, http://arxiv.org/abs/hep-th/0407266
Tegmark, M. 2004: Parallel Universes. In: Barrow, J. D., Davies, P. C. W., Harper, C.L. (eds.): Science and Ultimate Reality. Cambridge: Cambridge University Press, http://arXiv.org/abs/astro-ph/0302131
Tolman, R. C. 1934: Relativity, Thermodynamics and Cosmology. Clarendon Press: Oxford.

Turok, N., Hawking, S. W. 1998: Open Inflation, the Four Form and the Cosmological Constant. Phys. Lett. B432, 271-278, http://arxiv.org/abs/hep-th/9803156

Turok, N., Malcolm, P., Steinhardt, P. J. 2004: M Theory Model of a Big Crunch/Big Bang Transition, http://arxiv.org/abs/hep-th/0408083

Turok, N., Steinhardt, P. J. 2002: The Cyclic Universe: An Informal Introduction, http://arxiv.org/abs/astro-ph/0204479

Turok, N., Steinhardt, P. J. 2004a: Beyond Inflation: A Cyclic Universe Scenario, http://arxiv.org/abs/hep-th/0403020

Turok, N., Steinhardt, P. J. 2004b: The Cyclic Model Simplified, http://arxiv.org/abs/astro-ph/0404480

Vaas, R. 1994: Neue Wege in der Kosmologie. Naturwissenschaftliche Rundschau 47, 43-58.

Vaas, R. 1998: Is there a Darwinian Evolution of the Cosmos? Proceedings of the MicroCosmos – MacroCosmos Conference. Aachen, http://arXiv.org/abs/gr-qc/0205119

Vaas, R. 2000: Gibt das Anthropische Prinzip eine Antwort auf die Frage nach dem So-Sein unserer Welt? In: Fischbeck, H.-J. (Hrsg.): Warum ist die Welt so, wie sie ist? Mülheim an der Ruhr: Evangelische Akademie Mülheim an der Ruhr, 51-74.

Vaas, R. 2001a: Ewiges Leben im Universum? bild der wissenschaft 9, 62-67.

Vaas, R. 2001b: Vor dem Urknall – Modell Klassik. bild der wissenschaft 12, 43-49.

Vaas, R. 2001c: Vor dem Urknall – Modell Avantgarde. bild der wissenschaft 12, 52-55.

Vaas, R. 2001d: Quantenvakuum, Erbse oder Trivialität? bild der wissenschaft 12, 56-60.

Vaas, R. 2001e: Why Quantum Correlates Of Consciousness Are Fine, But Not Enough. Informação e Cognição 3, http://www.marilia.unesp.br/atividades/extensao/revista/v3/artigo4.html

Vaas, R. 2002a: Hawking & Co. bild der wissenschaft 5, 52-58.

Vaas, R. 2002b: Ewige Wiederkehr. bild der wissenschaft 5, 59-63.

Vaas, R. 2002c: Wenn die Zeit rückwärts läuft. bild der wissenschaft 12, 46-55.

Vaas, R. 2003a: Die Zeit vor dem Urknall. bild der wissenschaft 4, 60-67.

Vaas, R. 2003b: Das erste Licht. bild der wissenschaft 12, 44-51.

Vaas, R. 2003c: Der kosmische Code. bild der wissenschaft 12, 40-46.

Vaas, R. 2003d: Jenseits von Raum und Zeit. bild der wissenschaft 4, 50-56. English translation: Beyond Space And Time, http://arxiv.org/abs/physics/0401128

Vaas, R. 2003e: Problems of Cosmological Darwinian Selection and the Origin of Habitable Universes. In: Shaver, P. A., DiLella, L., Giménez, A. (eds.): Astronomy, Cosmology and Fundamental Physics. Berlin, Heidelberg, New York: Springer, 485-486.

Vaas, R. 2004a: Das Duell: Strings gegen Schleifen. bild der wissenschaft 4, 44-49. Extended English translation: The Duel: Strings versus Loops, http://arxiv.org/abs/physics/0403112

Vaas, R. 2004b: Der umgestülpte Urknall. bild der wissenschaft 4, 50-55. English translation: The Inverted Big-Bang, http://arxiv.org/abs/physics/0407071

Vaas, R. 2004c: Einstein und die Quantenwelt. bild der wissenschaft 8, 38-53.

Vaas, R. 2004d: Wie sich das Universum selbst schuf. bild der wissenschaft 10, 42-46.

Vaas, R. 2004e: Ein Universum nach Maß? In: Hübner, J., Stamatescu, I.-O., Weber, D. (Hrsg.): Theologie und Kosmologie. Mohr-Siebeck: Tübingen.

Vaihinger, H. 1911: Die Philosophie des Als Ob. Leipzig: Meiner 1927.

Veneziano, G. 1999: Challenging the Big Bang: a longer history of time. CERN Courier 39 (2), 18-20.

Vilenkin, A. 1982: Creation of universes from nothing. Phys. Lett. B117, 25-28.

Vilenkin, A. 1984: Quantum creation of universes. Phys. Rev. D30, 509-511.

Vilenkin, A. 2000: Eternal inflation and the present universe. Nucl. Phys. Proc. Suppl. 88, 67-74, http://arxiv.org/abs/gr-qc/9911087

Weinberg, S. 1993: Dreams of a Finat Theory. New York: Pantheon.

Wheeler, J. A., Zurek, W. H. (eds.) 1983: Quantum Theory and Measurement. Princeton: Princeton University Press.

Wike, V. S. 1982: Kant's Antinomies of Reason. Washington: University Press of America.

Wilkerson, T. E. 1976: Kant's Critique of Pure Reason. Oxford: Clarendon Press.

Wolfram, S. 2002: A New Kind of Science. Champaign: Wolfram Media.

Zeh, H. D. 2001: The Physical Basis of The Direction of Time. Berlin u. a.: Springer, http://www.time-direction.de/

# Acknowledgments

Though they are of course not responsible for the thoughts and proposals I offered here, I'd like to take this opportunity to thank Abhay Ashtekar, Hans-Joachim Blome, Martin Bojowald, Maurizio Gasperini, John Richard Gott III, Serdar Günes, Alan Guth, Jim Hartle, Bernulf Kanitscheider, Claus Kiefer, Li-Xin Li, Andrei Linde, Wolfgang Priester, Eckhard Rebhan, Carlo Rovelli, Urs Schreiber, Lawrence Schulman, Lee Smolin, Paul Steinhardt, Thomas Thiemann, Hakan Turan, Neil Turok, Gabriele Veneziano, Alex Vilenkin, Christian Wenzel, H. Dieter Zeh, and Thomas Zoglauer for discussions, kindness, and patience. And, as ever, I am very grateful to André Spiegel for his valuable suggestions and improvements. Thanks also to Alex Raab for the support with the figures.